\documentclass[lettersize,journal]{IEEEtran}
\usepackage{amsmath,amsfonts}
\usepackage{algorithmic}
\usepackage{algorithm}
\usepackage{array}
\usepackage[caption=false,font=footnotesize,labelfont=rm,textfont=rm]{subfig}
\usepackage{textcomp}
\usepackage{stfloats}
\usepackage{url}
\usepackage{verbatim}
\usepackage{graphicx}
\usepackage{cite}
\usepackage{xcolor}
\usepackage{enumerate}
\usepackage{multirow}
\usepackage{graphicx}
\usepackage{flushend}
\usepackage[colorlinks=false]{hyperref}
\hypersetup{
    colorlinks,
    citecolor=blue,
    filecolor=blue,
    linkcolor=blue,
    urlcolor=black
}
\begin{document}

\title{A Joint Planning Model for Fixed and Mobile
Electric Vehicle Charging Stations Considering Flexible Capacity Strategy}

\author{Zhe Yu, Xue Hu,
        and~Qin~Wang,~\IEEEmembership{Senior Member,~IEEE}

\thanks{Z. Yu, X. Hu and Q. Wang are with the Department of Electrical and Electronic Engineering, The Hong Kong Polytechnic University, Hong Kong (e-mail: qin-ee.wang@polyu.edu.hk). }
\thanks{}
\vspace{-0.5cm} 
}

\markboth{}%
{Shell \MakeLowercase{\textit{Yu and Wang}}: A Joint Planning Model for Fixed and Mobile Electric Vehicle Charging Stations}


\maketitle

\begin{abstract}
The widespread adoption of electric vehicles (EVs) has significantly increased demand on both transportation and power systems, posing challenges to their stable operation. To support the growing need for EV charging, both fixed charging stations (FCSs) and mobile charging stations (MCSs) have been introduced, serving as key interfaces between the power grid and transportation network. Recognizing the importance of collaborative planning across these sectors, this paper presents a two-stage joint planning model for FCSs and MCSs, utilizing an improved alternating direction method of multipliers (ADMM) algorithm. The primary goal of the proposed model is to transform the potential negative impacts of large-scale EV integration into positive outcomes, thereby enhancing social welfare through collaboration among multiple stakeholders. In the first stage, we develop a framework for evaluating FCS locations, incorporating assessments of EV hosting capacity and voltage stability. The second stage introduces a joint planning model for FCSs and MCSs, aiming to minimize the overall social costs of the EV charging system while maintaining a reliable power supply. To solve the planning problem, we employ a combination of mixed-integer linear programming, queueing theory, and sequential quadratic programming. The improved ADMM algorithm couples the siting and sizing decisions consistently by introducing coupling constraints, and supports a distributed optimization framework that coordinates the interests of EV users, MCS operators, and distribution system operators. Additionally, a flexible capacity planning strategy that accounts for the multi-period development potential of EVCS is proposed. This approach helps reduce both the complexity and the investment required for FCS construction. Finally, a case study with comparative experiments demonstrates the effectiveness of the proposed models and solution methods.

\end{abstract}
%
%

\section*{Nomenclature}
\vspace{-0.2cm}
\addcontentsline{toc}{section}{Nomenclature}
\begin{IEEEdescription}[\IEEEusemathlabelsep\IEEEsetlabelwidth{$C_{m,o}^{t}$/$\overline{C_{m,o}}$}]
	\item[\textbf{\emph{Sets}:}]
    \item[$\mathcal{I}$]     	Set of EV charging demands' loactions  $i$.
    \item[$\mathcal{J}$]     	Set of candidate charging points $j$.
    \item[$\mathcal{T}$]    Set of time interval indices $t$ = $1, \ldots , T$.
    \item[$\mathcal{K}$]      Set of fixed charging stations' locations $k$. 
     \item[$\mathcal{G}$]      Set of mobile charging stations' locations $g$. 
       \item[$\mathcal{R}$]      Set of EV achieving charging service sequence $r$ within time period.

	\item[\textbf{\emph{Parameters and Variables  }:}]
     \item[$HC_{n}$]     	Hosting capacity of EV load at node $n$.
      \item[$VSF_n$]     	Voltage stability factor at node $n$.
   \item[$\underline{V_{n}}$/$\overline{V_{n}}$]  Lower/upper limit of voltage at $n$. 
      \item[$\underline{P_{EV,n}}$/$\overline{P_{EV,n}}$]     	Lower/upper limit of active power of EV load at $n$.
      \item[$\underline{Q_{EV,n}}$/$\overline{Q_{EV,n}}$]     	Lower/upper limit of reactive power of EV load at $n$.
  \item[$\underline{\varphi_{EV,n}^{p}}$/$\overline{\varphi_{EV,n}^{p}}$]     Lower/upper limit of phase-$p$ power factor.
   \item[$N_{EV}$]     	Maximum number of EVs can be charged.
\item[$D_{MCS}$]  Distance between adjacent charging points.
\item[$R_{s}$]  Service radius of charging stations.
\item[$d_{EV}$]  Average driving distance of  EVs.
\item[$c_{tc}$]  Time cost for driving an EV to an EVCS. 
\item[$v_{avg}^t$]  Average driving speed of EVs during $t$. 
\item[$\xi_i^t$]  EV charging demand at node $i$ during $t$.
\item[$\delta_{g}^{t}$]  Average utilization rate of EVCS at $g$.
\item[$\iota _{g}^t$]  Capacity of  charging facilities at $g$ during $t$.
\item[$\underline{\iota}$/$\overline{\iota}$]  The minimum/maximum capacity of charging facilities at each charging point.
\item[$\overline{\iota_{tot}}$]  The maximum capacity of total EVCSs.
\item[$n_{m,g}^{t}$]  Number of chargers of  EVCS  at $g$ during $t$.
\item[$T_{w,g}^{t}$/$\overline{T_w}$]  Average waiting time/waiting time limit of  EVCS at $g$ during $t$.
\item[$C_{m,o}^{t}$/$\overline{C_{m,o}}$]  Total operation cost/operation cost limit of  MCSs during period $t$.
\item[$c_{m,o}^{t}$]  The operation cost per MCS during period $t$.
\item[$C_{cons}^{k}$]  Annual  cost of constructing the FCS at $k$.
\item[$C_{om}^{k}$]  Total annual costs of operating the FCS at $k$.
\item[$c_{om}^{k}$]  Per unit capacity costs of annually operating and managing the FCS at node $k$.
\item[$C_{loss}$]  Annual power loss costs of FCSs.
\item[$R_{f}$]  Annual net revenue of FCSs.
 \item[$V_{f,k}$]     Voltage of each bus if an FCS at node k. 
  \item[$\underline{V_{f}}$/$\overline{V_{f}}$]     Lower/upper limit of voltage.
 \item[$I_{f,k}$/$\overline{I_{f}}$]     Current/current limit of each branch if an FCS at node k.  
\item[$S_{m,g}^{t}$]  Available capacity of MCSs at $g$  during $t$.
\item[$S_{f,k}$]  Available capacity of FCS at node $k$.
\item[$SoC_{t,dep}^{r}$]  State of charge on departure of $r^{th}$ EV during $t$.
\item[$SoC_{t,arr}^{r}$]  State of charge on arrival of $r^{th}$ EV during  $t$.
\item[$\underline{P_{f}}$/$\overline{P_{f}}$]  Lower/upper limit of consumed active power.
\item[$\underline{Q_{f}}$/$\overline{Q_{f}}$] Lower/upper limit of reactive power.
\item[$S_{p,k}^t$]  Power purchased from the grid of FCS at $k$.
\end{IEEEdescription}


\section{Introduction}
\label{sec:Intro}

\IEEEPARstart{A}{s} a cleaner mode of transportation with lower emissions and energy consumption, electric vehicles (EVs) have attracted global attention \cite{zhang2015integrated}. With the increasing penetration of EVs into the coupled transportation and power systems, the strategic planning of electric vehicle charging stations (EVCSs) has become essential. The growing demand for EV charging introduces considerable load pressure and negative impacts on power systems, such as voltage fluctuations, three-phase voltage unbalance, power loss, and harmonics. While various EV charging management strategies have been proposed to address these problems, tackling them during the planning stage has proven to be more effective. By considering the potential impacts of EV integration in the early stages and adopting appropriate EVCS planning strategies, the negative effects on power systems can be mitigated more efficiently than by relying only on real-time operational measures. Many studies have proposed various approaches to optimize planning configurations by considering different constraints, decision-making variables, and performance indices, while there is a lack of collaborative planning perspective that integrates both the transportation and power sectors. EVCS planning is fundamentally a facility siting and sizing problem within a coupled transportation-power network. If the existing infrastructure fails to provide adequate power supply and charging services, the rapid growth of EV adoption may hinder practical deployment and reduce user satisfaction due to limited driving ranges. Therefore, it is necessary to consider EVCS planning in the context of coupled transportation and power systems, and promote cross-sector collaboration. Such coordination can help leverage the complementary strengths of both sectors, reduce EV load pressure, mitigate uncertainties, and improve the service level of regional distribution networks. 
\vspace{-0.3cm}
\subsection{Literature review}

Many studies focus on the optimal planning of EVCSs from the power system perspective. It is typically formulated as an optimization model with decision variables such as location, quantity, and capacity of charging stations \cite{lam2014electric}. Objective functions often include construction and operational costs, network performance metrics, and user convenience. For example, \cite{liu2012optimal} identifies optimal EVCS locations using a two-step screening method, followed by a capacity optimization model based on a modified primal-dual interior point algorithm to minimize total cost.   \cite{ahmad2021enhanced} employs an improved chicken swarm algorithm to determine EVCS placement while considering voltage profiles, voltage stability index, average voltage deviation index, and power losses. \cite{ahmad2023placement} integrates photovoltaic (PV) systems into the planning model and applies a hybrid algorithm combining gray wolf optimization and particle swarm optimization to relieve grid stress.  \cite {zhou2022location} proposes a location optimization model using a genetic algorithm to minimize total social cost, including construction, electricity usage, and carbon dioxide emissions.  References \cite{liu2012optimal, ahmad2021enhanced, ahmad2023placement, zhou2022location}  generally neglect transportation system  constraints, which may limit their practical applicability under real-world traffic conditions. 
 
Some studies approach EVCS planning from the perspective of transportation facility planning \cite {meng2024research,pourvaziri2024planning,calvo2024optimal}, disregarding power system constraints and impacts on power grid operation. In recent years, transportation-enabled mobile charging solutions, such as mobile charging stations (MCSs) and mobile energy storage facilities, have gained increasing attention due to their high practical value. For instance,   \cite {beyazit2023electric}  minimizes the total operation cost of MCSs and ensures fairness between MCSs and FCSs through a time-of-use pricing mechanism.  \cite {afshar2022mobile} proposes an optimization framework to coordinate EV charging in a multi-charger system that includes both FCSs and MCSs.  \cite  {aktar2024scheduling}  develops a dispatch optimization algorithm based on mixed-integer linear programming to reduce the number of EVs that fail to charge at fixed stations. \cite{he2023coordinated} presents a collaborative planning model based on the analytical target cascading method to enable data exchange between the upper and lower levels of the system, while it overlooks power system constraints. Most existing studies on MCS concentrate on dispatch strategies, with limited integration into long-term planning frameworks for energy systems.

Therefore, EVCS planning within a coupled power-transportation network merits deeper investigation. For example, \cite {li2022robust} formulates a robust optimization model for EVCS siting alongside distributed generation units within an integrated road and power network, employing kernel density estimation to improve the over-conservatism of robust optimization. \cite{deb2021robust} identifies candidate EVCS locations using a multi-objective framework that considers distance, traffic flow, and grid stability, and then applies a hybrid algorithm combining chicken swarm optimization and teaching-learning-based optimization to determine the optimal placement. Regardless of efforts to reduce the load pressure induced by EVCSs, the additional charging demand will inevitably have a significant impact on the power system. While the aforementioned studies aim to reduce the EV load burden, few provide planning frameworks that can mitigate or even transform these negative effects into positive outcomes. 

Meanwhile, planning approaches that incorporate renewable energy sources have received increasing attention due to their potential to reduce load fluctuations and support grid stability. For example, \cite{li2022robust} investigates a microgrid integrating wind power,  PV systems, EVCSs, and energy storage systems, and addresses uncertainties in EV charging and renewable generation through a robust optimization framework.   \cite{shaaban2019joint} considers the joint planning of DG units and EVCSs in a microgrid as a multi-objective mixed-integer nonlinear programming problem and solves it using a nondominated sorting genetic algorithm. \cite{li2020energy} and \cite{sun2021multi} explore capacity design strategies in conjunction with power scheduling schemes. However, these works primarily focus on in-station energy management and short-term operations, lacking a comprehensive framework for long-term infrastructure planning. 
\vspace{-0.3cm}
\subsection{Contributions of this paper}
\begin{figure}[t]  
    \centering      \includegraphics[width=0.5\textwidth]{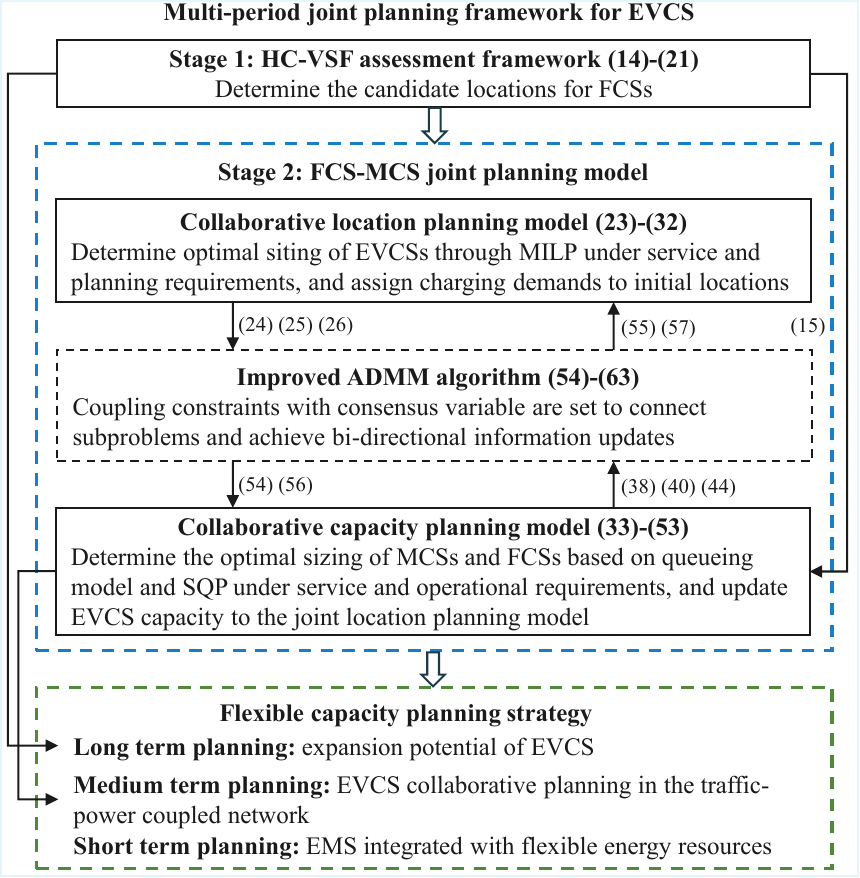}  
    \caption{Flowchart of the proposed multi-period EVCS joint planning framework.}  
    \label{flowchart}  
\end{figure}

This paper proposes a multi-period joint planning framework for EV charging facilities through cross-domain collaboration between the power and transportation sectors. The planning objective is to convert the negative impacts of EV integration into positive outcomes and improve social benefits. Derived from extensive EV charging, fixed charging stations (FCSs) and mobile charging stations (MCSs) serve as key interfaces in the coupled power and transportation networks.  Joint planning of FCSs and MCSs enhances the coupling depth and supports collaboration among multiple stakeholders to improve overall social welfare. In addition, current research mostly overlooks the long-term capacity development potential of EVCS, which is a key consideration of facility planning. In practical engineering, the construction of FCSs involves substantial investment, and grid expansion involves extensive engineering efforts. Given the foreseeable growth in EV adoption, it is of economic value to incorporate future expansion potential into the EVCS planning process. Therefore,  a multi-period collaborative planning framework is proposed in this paper. A flowchart is provided in Fig. \ref{flowchart}. The main contributions of this paper are summarized as follows:
\begin{itemize}
\item This paper proposes an ADMM-based two-stage joint planning model for FCSs and MCSs. In the first stage, a location evaluation framework incorporating EV hosting capacity (EV-HC) and voltage stability factor (VSF) assessments is established to determine candidate locations. In the second stage, an FCS-MCS joint planning model is developed. The optimal planning problem is solved by mixed-integer linear programming (MILP), queueing theory together with sequential quadratic programming (SQP). 
\item An improved alternating direction method of multipliers (ADMM) algorithm is adopted to couple the placement and capacity subproblems.  Coupling constraints are introduced to integrate global and local constraints through consensus variables.  A distributed optimization framework is established to coordinate the interests of different stakeholders: EV users, MCS operators, and distribution system operators (DSOs). 
\item Considering the multi-period capacity development potential of EVCS, a flexible capacity planning strategy is proposed to guide EVCS expansion over multiple periods. In the short-term planning,  an energy management system (EMS) integrated with multiple flexible energy resources is proposed to address operational problems. For medium-term planning, a collaborative capacity planning model is developed to reduce investment. For long-term planning, an evaluation framework for EVCS capacity expansion potential is developed to guide future development.
\end{itemize}

The rest of the paper is organized as follows. Section~\ref{sec:2} and Section \ref{sec:3} describe the first stage and the second stage of the ADMM-based FCS-MCS joint planning model in detail. Section~\ref{sec:4} presents the flexible capacity strategy for multi-period capacity planning. A case study on the proposed models is carried out in Section~\ref{sec:5}. Finally, Section~\ref{sec:6} concludes.

\section{Stage 1: Determination of candidate locations for FCSs}
\label{sec:2}
Based on Kirchhoff’s laws, the current and voltage relationships between the input (Node $n$) and the output (Node $m$) in distribution networks can be formulated as follows \cite{kersting2018distribution}: 
\vspace{-0.2cm}
\begin{equation}
\left[ I_{abc}^o \right]_{n} = \left[ I_{abc}^o \right]_{m} + \frac{1}{2} \left[Y_{abc} \right] \cdot \left[ V_{abc}^g \right]_{m}
\end{equation}
\vspace{-0.2cm}
\begin{equation}
\left[ V_{abc}^g \right]_{n} = \left[ V_{abc}^g \right]_{m} + \frac{1}{2} \left[ Z_{abc} \right] \cdot \left[ I_{abc}^o \right]_{m}
\end{equation}
where $\left[ I_{abc}^o \right]_{n}$ means the matrix of the three-phase currents flowing from node $n$ with loss removed$, \left[ V_{abc}^g \right]_{n}$ means the three-phase voltage to earth of node $n$, $\left[Y_{abc} \right]$ and $\left[ Z_{abc} \right]$ represent the impedance and admittance matrices between node $n$ and node $m$. (1) and (2) can be put into partitioned matrix form, $\left[U \right]$ means identity matrix. The basic distribution system line model can be developed as follows: 
\begin{equation}
    \begin{bmatrix}
        V_{abc}^g \\ I_{abc}^o
    \end{bmatrix}_{n} =
    \begin{bmatrix}
        [A] & [B] \\
        [C] & [D]
    \end{bmatrix} \cdot
    \begin{bmatrix}
        V_{abc}^g \\ I_{abc}^o
    \end{bmatrix}_{m}
\end{equation}
\begin{equation}
    [B] = [Z_{abc}] 
\end{equation}
\begin{equation}
    [C] = [Y _{abc}] + \frac{1}{4}  [Y _{abc}] \cdot [Z_{abc}] \cdot [Y _{abc}]
\end{equation}
\begin{equation}
    [A] = [D] = [U] + \frac{1}{2}  [Z_{abc}] \cdot [Y _{abc}]
\end{equation}
Regarding each phase $p=a,b,c$, the relationship between voltage and current of each phase $p$ at node $n$ can be calculated as (7) and (8) for simplification, $V_{p,n}$ and $I_{p,n}$  represent voltage and current of phase $p$ at node $n$,
\begin{equation}
V_{p,n} = V_{p,m}+Z_{pa} I_{a,m}+Z_{pb} I_{b,m}+Z_{pc} I_{c,m}
\end{equation}
\begin{equation}
V_{p,n} = V_{p,m}+(R_{p}+jX_p)I_{p,m}
\end{equation}
The active power $P_{p,n}$, reactive power $Q_{p,n}$, and apparent power $S_{p,n}$ can be achieved in the following formulas  \cite{modarresi2016comprehensive}:
\begin{equation}
P_{p,n}=\frac{V_{p,m} V_{p,n} \cos (\theta _{p,m}-\theta _{p,n})-V_{p,n}^{2}-Q_{p,n} X_p}{R_p} 
\end{equation}
\begin{equation}
Q_{p,n}=\frac{P_{p,n} X_p-V_{p,m} V_{p,n} \sin (\theta _{p,m}-\theta _{p,n})}{R_p}
\end{equation}
\begin{equation}
S_{p,n}=P_{p,n}^2+Q_{p,n}^2
\end{equation}
From (9) to (11), a fourth-degree equation can be derived:
\begin{equation}
V_{p,n}^{4}\!\!+\!\left(2 P_{p,n} R_{p}+2 Q_{p,n} X_{p}-V_{p,m}^{2}\right) V_{p,n}^{2}\!\!+\!S_{p,n}^{2} Z_p^{2}=0
\end{equation}
Considering the limited capacity under normal operation, the maximum load ability can be achieved under the condition that the discriminant of (12) is equal:
\begin{equation}
    S_{p,n}^{\max} = \frac{V_{p,m}^2}{4Z_p \cos^2 \left( \frac{\omega- \phi }{2} \right)}
\end{equation}
where $\omega=\tan^{-1} X/R, \phi=\tan^{-1} Q_n/P_n$.

Based on the above analysis, an assessment framework for FCS placement is developed through evaluating EV-HC and VSF of nodes in distribution networks. VSF quantifies the variation of node voltage with increasing active power. The node's maximum holding capacity is evaluated by EV-HC assessment. This framework aims to determine the initial locations of FCSs through identifying the maximum allowable EV load of each node and assessing the sensitivity level to EV integration without endangering system stability. Strong nodes with high EV-HC and low VSF can be selected as candidate locations for FCS deployment, while weak areas are flagged for reinforcement to mitigate potential negative impacts. The assessment framework can be formulated as follows:

 $VSF_{n}$ is defined as,
 \vspace{-0.3cm}
 \begin{equation}
VSF_{n}=\left |\frac{d V_{n}}{d P_{EV,n}}\right |
\end{equation}
\vspace{-0.3cm}

 $HC_{n}$ is defined as,
 \vspace{-0.3cm}
 \begin{equation}
HC_{n} = \sum_{l=0}^{N_{EV}} \sum_{p=a}^{c} P_{e,l}^{p,n}
\end{equation}
\vspace{-0.3cm}

subject to
\begin{equation}
\underline{V_{n}} \leq V_{n} \leq \overline{V_{n}}
\end{equation}
\begin{equation}
    \left| \frac{|V_{n}^{p}| - V_{n}^{avg}}{V_{n}^{avg}} \right| \leq \gamma 
\end{equation}
\begin{equation}
    \left| I_{mn} \right| \leq \overline{I_{mn}} 
\end{equation}
\begin{equation}
\underline{P_{EV,n} }\leq P_{EV,n}\leq\overline{P_{EV,n} }
\end{equation}
\begin{equation}
\underline{Q_{EV,n} }\leq Q_{EV,n}\leq\overline{Q_{EV,n} }
\end{equation}
\begin{equation}
    \underline{\varphi_{EV,n}} \leq \varphi_{EV,n}^{p} \leq \overline{\varphi_{EV,n}}
\end{equation}
where $P_{e,l}^{p,n}$ represents the total EV charging power with EV quantity $l$ of phase $p$ at node $n$. $V_{n}$, $V_{n}^p$ and  $V_{n}^{avg}$ represent voltage at node $n$, phase-$p$ voltage at node $n$ and average voltage across three phases, respectively. $I_{mn}$ and  $\overline{I_{mn}}$ represent the current and thermal limit in branch from node $m$ to node $n$. $P_{EV,n}$ and $Q_{EV,n}$ mean total active power and reactive power of EV load at node $n$. $\varphi_{EV,n}^{p}$ is phase-$p$ power factor for EV load at node $n$. (16)-(18) present the voltage limits, three-phase voltage unbalance limits and current limits, respectively. (19)-(21) ensure EV charging power and power factor at the charging point within agreed ranges. 

\section{Stage 2: FCS-MCS joint planning model}
\label{sec:3}

According to the candidate locations of FCSs in Stage 1, the collaborative placement and capacity of FCSs and MCSs are planned based on the joint model in Stage 2. The joint planning of charging stations needs to consider the interests of multiple stakeholders. The objective function of the proposed model is defined as follows:
\vspace{-0.1cm}
\begin{equation}
    \text{Minimize} \quad  F=C_{trav} +C_{MCS}+C_{FCS}
\end{equation}
\noindent where $C_{trav}$ is the driving time cost of EV users, $C_{MCS}$ is the cost of operating the MCSs, and $C_{FCS}$ is the annual net cost of DSO.
\vspace{-0.4cm}
\subsection{Collaborative location planning}

Based on the determined locations of FCSs in the first stage, the optimal placements of MCSs are subsequently determined in order to minimize the total time-related cost incurred by driving EVs to the charging stations.

Consider a transportation network comprising $N$ candidate nodes, where certain nodes can be selected as MCS locations. The decision variables, station selection variables and location assignment variables are defined as: set ${x_j^t}$ as a binary decision variable indicating whether node $j$ is selected to place EVCS during period $t$, ${x_j^t} = 1$ if MCSs are placed at node ${j}$, and ${x_j^t} = 0$ otherwise. ${y_{ij}^t}$ is set to indicate whether EV charging demand at $i$ can be served by EVCS at ${j}$, ${y_{ij}^t}=1$  if demand at $i$ can be served by EVCS at ${j}$, and ${y_{ij}^t} = 0$ otherwise. ${v_{ij}^t}$ is set to indicate whether EV charging demand at $i$ is scheduled to EVCS at ${j}$, ${v_{ij}^t} = 1$ if demand at $i$ is scheduled to $j$, and ${v_{ij}^t} = 0$ otherwise. 

EV users who require charging services need to drive EVs to the charging points, which incurs driving distance and related time cost. In this study, the distance matrix $D_{ij}$ represents the shortest routes between demand points and charging points, which is computed by the Floyd-Warshall algorithm. The objective of collaborative location optimization is to minimize the total time cost for all EVs to reach charging stations under the constraints of service and planning requirements. The mathematical model can be formulated as follows:
\begin{align}
    \text{Minimize} \quad & C_{trav} = \sum_{j \in J}\sum_{i \in I} \sum_{t \in T} \frac{c_{tc}\xi _i^t y_{ij}^t D_{ij}^t}{v_{avg}^t}   \\[-2pt] 
    \text{subject to} \quad & x_j^{t} \geq y_{ij}^t, \quad \forall i \in I, j \in J  \\[-2pt]
    & \sum_{j \in J } x_j^t = \psi  , \quad \forall j \in J  \\[-2pt]
    &  x_j^{t}y_{ij}^t \geq v_{ij}^t, \quad \forall i \in I,j \in J \\[-2pt]
    & \sum_{j \in J} v_{ij}^t = 1, \quad \forall t \in T , i \in I \\[-2pt]
    &R_s \leq  D_{MCS} \leq d_{EV}  \\[-2pt]
    & x_k=1, \quad \forall k \in K  \\[-2pt]
    &  D_{ij}^{t}y_{ij}^t \leq \varsigma d_{EV}, \quad \forall t \in T , i \in I, j \in J   \\[-2pt]
    &c_{tc} = \frac{M_a}{t_{wa}} \\[-2pt] 
    & x_{ij}^t, y_{ij}^t, v_{ij}^t \!\in \!\{0,1\}, \quad \forall t \in T , i \in I, j \in J
\end{align}

\noindent where (24) and (25) are service availability constraints and quantity constraints of charging points, respectively.  (26) means that only when the node $j$ has been placed with charging facilities and its scope of service can cover the demand at node $i$, which can be arranged to point  $j$ during period $t$. (27) ensures all EVs in every time period have access to charging facilities. (28) shows the distance requirement between adjacent charging points.  (29) sets the location for FCS. (30) is the constraints of EV driving distance, $\varsigma$ means user psychological factor. (31) presents driving time cost from starting place to charging place, $M_a$ represents the average annual income of residents in this area, $t_{wa}$ represents the average annual working hours of residents in this area. (32) are binary restrictions.
\vspace{-0.4cm}
\subsection{Collaborative capacity planning}
\subsubsection{MCS capacity}
Regarding capacity planning of MCSs, it is essential to consider the utilization efficiency of charging equipment and user queueing time to reduce ineffective operational costs and improve user satisfaction. Therefore, the primary objective of MCS capacity optimization is to minimize the total cost of operating MCSs while reducing the user waiting time cost. To capture the dynamic charging demands, M/M/c queueing model is integrated into the MCS capacity optimization framework. The model of MCS capacity optimization can be formulated as follows: 
\vspace{-0.1cm}
\begin{align}
    \text{Minimize} \quad & C_{\text{MCS}} = \sum_{t \in T} \sum_{g \in G} \left( C_{m,o}^t + c_{tc}  \lambda_g^t  T_{w,g}^t \right)  \\[-2pt]
    \text{subject to} \quad
    & \rho_g^t = \frac{\lambda_g^t}{n_{m,g}^t \mu_g^t} < 1 , \quad \forall g \in G  \\[-2pt]
    & P_{0,g}^t\!\! = \!\!\!\left[\sum_{l=0}^{n_{m,g}^t\!-1} \!\!\frac{(\lambda_g^t/\mu_g^t)^l}{l!} \!+\! \frac{(\lambda_g^t/\mu_g^t)^{n_{m,g}^t}}{n_{m,g}^t! (1 \!- \!\delta_g^t)} \!\right]^{-1}  \\[-2pt]
    & L_{q,g}^t = \frac{(n_{m,g}^t \delta_g^t)^{n_{m,g}^t}  \delta_g^t}{n_{m,g}^t! (1 - \delta _g^t)^2}  P_{0,g}^t  , \quad \forall g \in G  \\[-2pt]
    & L_{s,g}^t = L_{q,g}^t + \frac{\lambda_g^t}{\mu_g^t}  , \quad \forall g \in G  \\[-2pt]
    &\underline{\iota}\leq \iota_g^t \leq  \overline{ \iota}  , \quad \forall g \in G  \\[-2pt]
    & T_{w,g}^t = \frac{L_{q,g}^t}{\lambda_g^t} \leq \overline{T_w} , \quad \forall g \in G   \\[-2pt]
    & C_{m,o}^t = \sum_{g \in G} c_{m,o}^t n_{mcs}^t \leq \overline{C_{m,o}} 
\end{align}
\noindent where (34) defines the constraints of maintaining steady state at charging points \textit{g} in time period  \textit{t} through the utilization law  $\rho_g$. $\lambda_{g}^{t}$ and $\mu_{g}^{t}$ mean the arrival rate and service rate of  EVCS at $g$ during $t$. In(35), $P_{0,g}^{t}$ represents the state probability of no EV receiving charging services in the charging point \textit{g} during time period  $t$. (36) describes the average queue length $L_{q,g}^t$, and (37)  describes the total number of vehicles in the system $L_{s,g}^t$, including those in service and in the queue. (38) presents the capacity limits for each charging point $g$ during period $t$. (39) gives the average waiting time $T_{w,g}^t$, obtained by Little’s Law. (40) is the constraints of operation cost of MCSs, $n_{mcs}^t$ means the total number of MCSs employed during period $t$.
 
\subsubsection{FCS capacity}
Regarding capacity planning of FCSs, the objective function is defined as the minimization of the annual net cost of FCSs, which includes construction cost, operation cost, and network power loss cost, while deducting the net revenue from EV charging service. The mathematical model can be formulated as:
\vspace{-0.1cm}
\begin{align}
\text{Minimize} \quad & C_{FCS} = \!\sum_{k\in K} (C_{cons}^{k} \!+ \!C_{om}^{k})\! + C_{loss} \!\!- R_{f}  \\[-2pt]
\text{subject to} \quad
& \underline{V_{f}} \leq V_{f,k} \leq \overline{V_{f}}, \quad \forall k \in K    \\[-2pt]
& \left |I_{f,k}  \right | \leq \overline{I_{f}} ,\quad \forall k \in K  \\[-2pt]
& \sum_{g\in G }\!S_{m,g}^t\!\!+\!\!\!\sum_{k\in K } \!S_{f,k} \!\!\geq \!\!\!\sum_{r\in R} \!SoC_{t,dep}^{r}\!\!\!-\!\!SoC_{t,arr}^{r} \\[-2pt]
& \underline{P_f} \leq \sum_{k\in K} P_{f,k} \leq \overline{P_f}  \\[-2pt]
& \underline{Q_f} \leq \sum_{k\in K} Q_{f,k} \leq\overline{Q_f}  \\[-2pt]
& S_{q,k}\leq S_{f,k}\leq HC_k  ,\quad \forall k \in K \\[-2pt]
& (P_{p,k}^t)^2 + (Q_{p,k}^t)^2 = S_{p,k}^t \leq \sigma S_{f,k} 
\end{align}
\noindent where (42) and (43) represent voltage limits and the permitted maximal current limits in each feeder, respectively. (44) means the total electricity supply capacity of FCSs and MCSs in time period $t$ should cover the charging demands, which reflects the supply capability of the coupled traffic-power system. (45)-(46) show the active and reactive power limits.  $P_{f,k}$ and $Q_{f,k}$ mean consumed active and reactive power of FCS at $k$. (47) is the restriction of FCS capacity, the upper limit is based on EV-HC assessment, and the lower limit value $S_{q,k}$ is calculated from queuing theory.  (48) indicates that the power purchased from the main grid should not exceed the capacity of the charging station. 

\textit{(1) Construction cost}
\vspace{-0.2cm}
\begin{equation}
\begin{split}
C_{cons}^{k} &= \text{CRF}  (C_{base}^{k}+C_{inve}^{k} S_{f,k}) \\
&=\frac{h(1+h)^\varepsilon }{(1+h)^\varepsilon  - 1}  (C_{base}^{k}+(C_{char}^{k} \\&+ C_{ener}^{k}+ C_{land}^{k}) S_{f,k})
\end{split}
\end{equation}
\noindent where CRF means capital recovery factor, $h$ is the discount rate, $\varepsilon $ is the depreciation period. $C_{cons}^k$ is the construction cost of FCS at point $k$. $C_{base}^{k}$ is the basic infrastructure cost for an FCS. $C_{inve}^{k}$ represents the per-unit capacity investment cost of the FCS at $k$, where $C_{char}^{k}$, $C_{ener}^{k}$, $C_{land}^{k}$ are the per-unit capacity investment cost of charging devices and charger supporting facilities, energy coordination devices, such as photovoltaic panels and batteries, and the cost of land, respectively.
 
\textit{(2) Operation cost, power loss cost and charging station revenue}
\vspace{-0.3cm}
\begin{equation}
\begin{split}
C_{om}^{k} =(c_{oper}^{k}+c_{mant}^{k}+c_{hr}^{k}+  c_{vg}^{k}) S_{f,k}
\end{split}
\end{equation}
\begin{equation}
C_{loss}=  t_{om}  c_{pb}P_{loss}  
\end{equation}
The operation cost of the FCS at node $k$ includes daily basic operating cost $c_{oper}^{k}$, device management and maintenance cost $c_{mant}^{k}$, human resource cost $c_{hr}^{k}$, and V2G management cost $c_{vg}^{k}$. The power loss cost of distribution networks with the newly planned EV charging stations can be formulated in (51), where $P_{loss}$ is the power loss of the distribution system, $t_{om}$ is the annual utilization hours.
\vspace{-0.1cm}
\begin{equation}
R_f =\sum_{k\in K}  t_{om}S_{f,k} (c_{ps}+c_{cs}-c_{pb}-c_{tax}) -\sum_{k\in K}  c_{bf}S_{f,k}
\end{equation}
\vspace{-0.1cm}
\begin{equation}
c_{tax}=R_{xp} (c_{ps}-c_{pb})+R_{xs}c_{cs}
\end{equation}
 \noindent where $R_f$ is the annual net revenue from operating FCSs. $c_{pb}$, $c_{cs}$ and $c_{ps}$ mean the price of electricity purchased from the grid, service price, and selling price for EV charging, respectively. $c_{bf}$ is the base electricity price for high-demand industrial and commercial electricity consumption according to the maximum power demand. $R_{xp}$ and $R_{xs}$ are the electricity tax rate and service tax rate, which are 13\% and 6\% in this study, respectively.  It should be mentioned that each cost item mentioned in this work has been converted into the cost for each year of the utilization period. 
\vspace{-0.3cm}
\subsection{Improved Alternating Direction Method of Multipliers Algorithm}

To address the siting-sizing coupling planning problem while considering the different ownership structures of the stakeholders, this paper employs an improved ADMM algorithm to solve the FCS-MCS joint planning model. ADMM is an algorithm designed to solve convex joint optimization problems by decomposing them into smaller, more manageable subproblems \cite{xuan2025admm}. In distributed optimization, each stakeholder independently pursues its own profit maximization, while coordination is achieved through the coupling variables provided by other parties. This negotiation process is simulated by the iterations of ADMM, with the negotiation objective corresponding to the convergence condition of ADMM. In addition, due to the inclusion of both binary siting decisions \( x_j \in \{0,1\} \)  and integer capacity decisions \( z_j \), the problem introduces non-convex coupling relationships, making the convergence of primal residual difficult. This paper introduces coupling constraints with consensus variable \( w_j \) to effectively resolve the aforementioned issues and enhance convergence speed.

\subsubsection{Coupling constraints} Coupling constraints are designed to link the siting and sizing decisions.  Specifically,  local constraints are introduced in the sitting subproblem, global capacity constraints are incorporated into the capacity subproblem, and the consensus constraint \( W \) is adopted to connect the location block \( X \) and the capacity block \( Z \). The coupling constraints are formulated as:  
\vspace{-0.2cm}
\begin{equation}
   w_j \leq \overline{\iota} \  x_j, \quad \forall j \in J
  \end{equation}
\begin{equation}
    \sum_{j\in J} z_j \leq \overline{\iota _{tot}}
  \end{equation}
  \begin{equation}
    w_j = z_j, \quad \forall j\in J
\end{equation}
In the siting subproblem, \( X \) and \( W \) are coupled through linearized constraints.  In the capacity subproblem, global capacity constraints are added, and the consistency constraint \( w_j = z_j \) is achieved. A projection operator is used to update \( w_j \), ensuring \( w_j \) is within the feasible range. The projection method is defined as:
\vspace{-0.2cm}
\begin{equation}
  w_j^{k+1} = \min \left( \max(z_j^{k+1} + u_j^{k}, 0), \overline{\iota} x_j^{k+1} \right)
\end{equation}

\subsubsection{Promoting primal residual convergence} To promote convergence of the primal residual during the ADMM iterations, this model applies the Big-M method to linearize binary siting variables. This ensures that the capacity of unselected nodes is zero, while \( w_j \) can vary within \([0, \overline{\iota} ]\) when \( x_j = 1 \). In addition, a penalty term \( \frac{ \rho}{2} \| z^k - w^k \| \) is added to the objective function, and Lagrangian multiplier terms are included for violations of the capacity constraints. This ensures that during alternating optimization, the update of \( X \) is influenced by \( W \), and the update of \( Z \) strives to follow \( W \), leading to consistency between \( X \) and \( Z \) in terms of effective capacity.

\subsubsection{Solution of subproblems} The siting subproblem is formulated as a mixed-integer linear programming model to handle discrete variables. In the joint sizing problem, the optimal capacity of MCSs is solved using queueing theory and bisection method. If the result exceeds limits, a proportional scaling method is applied for adjustment. Based on the derived lower and upper bounds for the FCS capacity, MATPOWER is used to solve the optimal capacity of FCS.

\subsubsection{ADMM Update Rules}
The updates in the ADMM framework proceed as follows:

\begin{enumerate}[(i)]
   \item {EVCS placement:}
\vspace{-0.2cm}
  \begin{equation}
    x^{k+1} = \arg\min \left( f(x) + \frac{ \rho}{2}\| z^k - w^k + u^k \| \right)
  \end{equation}  
  \item {EVCS capacity:}
\vspace{-0.2cm}
  \begin{equation}
    z^{k+1} = \arg\min g(z) 
  \end{equation} 
  \item {Consensus variable:}
\vspace{-0.2cm}
  \begin{equation}
    w^{k+1} = \Pi_{0\leq w \leq \overline{\iota}x^{k+1}} \left( z^{k+1} + u^k \right) 
  \end{equation} 
  \item {Dual variable:}
\vspace{-0.2cm}
  \begin{equation}
    u^{k+1} = u^k + \rho(z^{k+1} - w^{k+1})
  \end{equation}
  \item Convergence and stopping criteria:
\end{enumerate}
\begin{equation}
  \| r_{\text{prim}} \| = \| z^{k+1} - w^{k+1} \| \leq \epsilon_{\text{prim}} 
  \end{equation}
  \begin{equation}
  \| r_{\text{dual}} \| = \rho \| w^{k+1} - w^k \| \leq \epsilon_{\text{dual}}
\end{equation}

This ADMM-based joint model can effectively coordinate the discrete siting problem and continuous capacity decisions to harmonize the interests of EV users, MCS operators, and the DSO.  Each subproblem can be independently updated through its own optimization method, while information is exchanged through a consensus variable. In addition, the parallel processing architecture of this model has decomposability and scalability. Task-specific objective functions and application scenarios can be achieved by adding relevant requirements in subproblems under the framework.
\vspace{-0.2cm}
\section{Flexible capacity strategy}
\label{sec:4}

In this paper, considering multi-period facility planning, a flexible capacity planning strategy is proposed to approach EV charging demand evolution. The planning horizon is divided into three timeframes: (i) long-term capacity expansion potential at the original sites, (ii) medium-term flexible capacity planning from collaborating FCSs and MCSs, and (iii) short-term operational strategy integrating multiple flexible energy resources. The flexible capacity strategy is presented in Table \ref{flexible capacity strategy}. The long-term and medium-term planning models have been described previously, while the short-term energy management strategy is detailed in the following part.
\renewcommand {\arraystretch} {1.5}
\begin{table*}[]
\caption{Flexible capacity strategy}
\label{flexible capacity strategy}
\vspace{-0.2cm}
\begin{tabular}{ccll}  \hline
Time period &
  Two-stage model&
  \multicolumn{1}{c}{Technical fundamental} &
  \multicolumn{1}{c}{Approach to charging demand increase} \\  \hline
Future 20 years &
  Stage 1 &
  HC-VSF assessment framework &
  \multicolumn{1}{c}{\begin{tabular}[c]{@{}c@{}}The expansion potential of EVCS at original site\end{tabular}} \\[-1ex]
Future 10 years &
  Stage 2 &
  FCS-MCS joint planning model&
  \multicolumn{1}{c}{\begin{tabular}[c]{@{}c@{}}Capacity planning of new facilities  based on coupled traffic-power  networks\end{tabular} } \\[-1ex]
Short-term operation &
  - &
  Energy management strategy &
  \multicolumn{1}{c}{\begin{tabular}[c]{@{}c@{}}Real-time flexible energy coordination\end{tabular}} \\  \hline \end{tabular}
\end{table*}

\begin{figure}[t]  
    \centering      \includegraphics[width=0.5\textwidth]{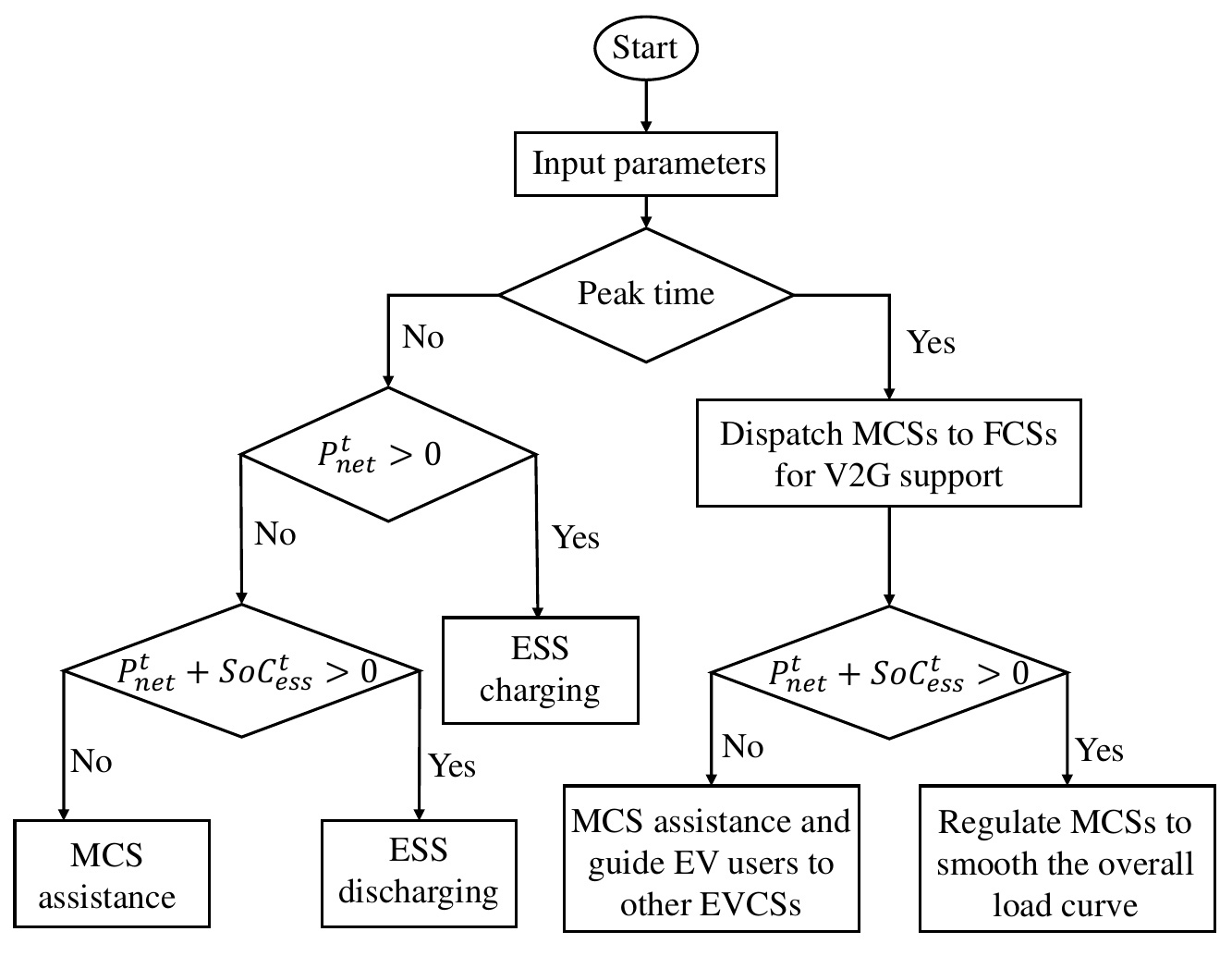}  
    \caption{Flowchart of the proposed EMS.}  
    \label{ems1}  
\end{figure}

The proposed EMS integrates multiple flexible energy resources, including photovoltaic (PV) generation, energy storage systems (ESSs), vehicle-to-grid (V2G) services from EVs and MCSs. To alleviate peak load in the distribution system, this strategy enables real-time dispatch of MCSs to support FCSs via V2G. The net power in the grid ${P}_{net}^t$, defined as the difference between power supply and demand, is given by: 
\vspace{-0.2cm}
\begin{equation}
\begin{split}
    {P}_{net}^t ={P}_{grid}^t+{P}_{pv}^t+{P}_{mcs}^t+{P}_{vg}^t-{P}_{ev}^t-{P}_{load}^t
\end{split}
\end{equation}
where $P_{grid}^t$,  $P_{pv}^t$, $P_{mcs}^t$ mean power supplied by the main grid,  power generated by PV and power delivered by MCSs during period $t$, respectively. $P_{vg}^t$ represents V2G power from EVs, and $P_{ev}^t$ means power consumed by EV charging during $t$. When net power ${P}_{net}^t $ is positive, the ESS is charged. Otherwise, the ESS switches to discharging mode to compensate for the shortfall.  In the valley and flat periods, V2G support from MCSs is utilized when ESS capacity is insufficient. The operational process of EMS is illustrated in Fig. \ref {ems1}.
\vspace{-0.2cm}
\section{Case Studies}
\label{sec:5}
\subsection{ Case Overview and Parameter Settings }
\begin{figure}[t]  
    \centering      \includegraphics[width=0.28\textwidth]{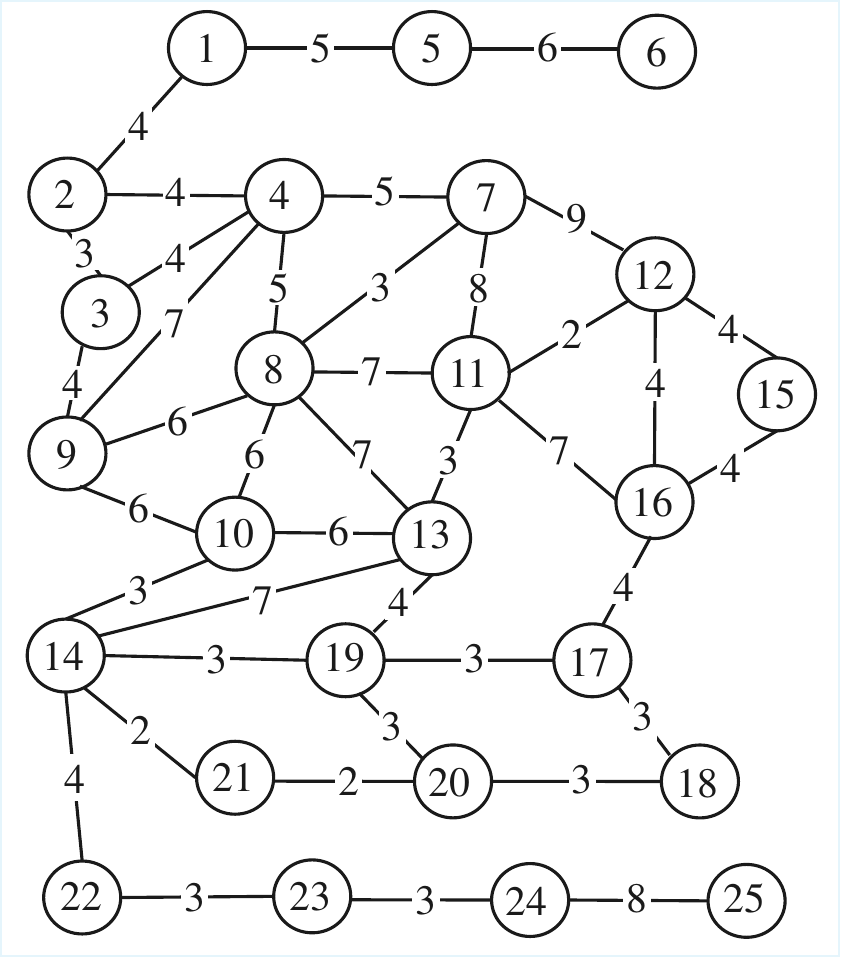}  
    \caption{The 25-node traffic network diagram used for the case study.}  
    \label{25traffic}  
\end{figure}

\begin{figure}[h]  
    \centering      \includegraphics[width=0.35\textwidth]{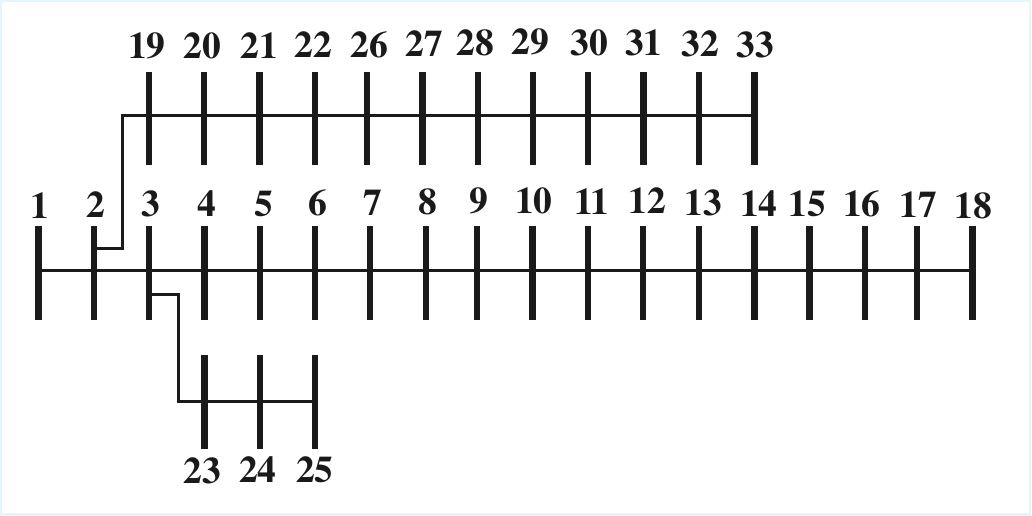}  
    \caption{The IEEE 33-node power network diagram used for the case study.}  
    \label{33ieee}  
\end{figure}

The calculation examples and comparison tests are conducted using real-world  EV charging data from Shenzhen, China \cite{li2025urbanev}. A coupled system comprising an IEEE-33 node power system \cite{ahmad2023placement} and a 25-node transportation system \cite{wang2013traffic} is adopted to illustrate the proposed planning method. The networks of the 25-node traffic system and the 33-node power system are depicted in Fig. \ref{25traffic} and Fig. \ref{33ieee}, and the coupling points between the two networks are shown in Table \ref{couplenodes}. In Fig. \ref{33ieee}, the number on each arc represents the distance between the corresponding two nodes with a per-unit distance of 10 km. The 33-node power system is configured with real load data, with load type details provided in Table \ref{loadtype}.  In this study, the MCSs equipped with a 600 kWh modular battery pack and 4 chargers are chosen for case tests, each supporting a charging rate of up to 100 kW. These MCSs are integrated with solar panels to enable self-charging of the batteries. The construction-related costs and operation-related costs are referenced from \cite{zhang2020mobile}. The EV charging electricity prices, service fees, and on-grid electricity tariffs are obtained from \cite{ref_powerprice}. The travel and queuing time costs for EV users are calculated based on \cite{ref_incomeshenzhen2023}. The parameter settings in this paper are presented in Table \ref{parameters}. The proposed models are modeled on MATLAB 2024b on a laptop with an i7 2.40 GHz processor and 16GB RAM. 

\renewcommand {\arraystretch} {1.5}
\begin{table}
\caption{Parameters of case study}
\vspace{-0.2cm}
    \centering
    \begin{tabular}{cccccc}\hline
         Para.&  Value&  Para.&  Value&  Para.& Value\\\hline
         $\overline{I_{mn}} $ &  1000 A&  $\overline{T_{w}} $&  1/6 h&  $\overline{C_{m,o}}$&  90 \$/h\\[-0.5ex]
         $\varphi_{EV,n}^{p}$&  0.95&   $c_{tc}$&  8.15 \$/h&  $\underline{V_{f}} $& 0.9 p.u.\\[-0.5ex]
         $\gamma$ &  0.03&   $\mu_j$&  4&  $\overline{V_{f}}$& 1.1 p.u.\\[-0.5ex]
         $v_{aver}^t$&  40 km/h&  $\underline{\iota }$&  0&  $r$& 0.07\\[-0.5ex]
         $\psi $&  5&   $\overline{\iota }$&  20&  $n$& 10\\[-0.5ex]
         $R_s$ &  10 km&  $c_{m,o}^t$&  0.9 \$/h&  $c_{om}$& 32.9 \$/kW\\[-0.5ex]
         $d_{EV}$&  100 km&  $C_{base}^{k}$&  288,000 \$&   $C_{land}^k$&  0.014 \$/kW\\[-0.5ex]
 $\varsigma$& 0.3& $C_{inve}^k$& 197 \$/kW& $t_{om}$ & 8 h/day\\[-0.5ex]
  $c_{pb}$& 0.078 \$/kWh& $c_{ps}$&  0.13 \$/kWh& $c_{cs}$& 0.11 \$/kWh\\[-0.5ex]
 $c_{bf}$&  69.8 \$/kWh& & & &\\\hline
    \end{tabular}
    
    \label{parameters}
\end{table}

\vspace{-0.2cm}
\vspace{-0.2cm}

\renewcommand {\arraystretch} {1.5}
\begin{table}[]
\centering
\caption{Coupling nodes between the 25-node traffic network and the 33-node power network}
\vspace{-0.2cm}
\label{couplenodes}
\begin{tabular}{>{\centering\arraybackslash}p{0.45\linewidth}>{\centering\arraybackslash}p{0.05\linewidth}>{\centering\arraybackslash}p{0.05\linewidth}>{\centering\arraybackslash}p{0.05\linewidth}>{\centering\arraybackslash}p{0.05\linewidth}>{\centering\arraybackslash}p{0.05\linewidth}} \hline 
Coupling Node ID       & \multicolumn{1}{c}{01} & \multicolumn{1}{c}{02} & \multicolumn{1}{c}{03} & \multicolumn{1}{c}{04} & \multicolumn{1}{c}{05} \\[-1ex]
Distribution Node ID   & \multicolumn{1}{c}{2}  & \multicolumn{1}{c}{19} & \multicolumn{1}{c}{20} & \multicolumn{1}{c}{22} & \multicolumn{1}{c}{3}  \\[-1ex]
Transportation Node ID & \multicolumn{1}{c}{1}  & \multicolumn{1}{c}{5}  & \multicolumn{1}{c}{11} & \multicolumn{1}{c}{16} & \multicolumn{1}{c}{4}  \\ \hline 
Coupling Node ID       & 06 & 07 & 08 & 09 & 10 \\[-1ex]
Distribution Node ID   & 15 & 23 & 25 & 7  & 18 \\[-1ex]
Transportation Node ID & 20 & 23 & 25 & 8  & 18 \\ \hline 
\end{tabular}
\end{table}
 \vspace{-0.2cm}
\begin{table}[]
\centering
\setlength{\tabcolsep}{4mm}{
\caption{Load types in the 33-node power system}
\vspace{-0.2cm}
\label{loadtype}
\begin{tabular}{clllll} \hline  
\multicolumn{1}{c}{Load Type}                    & \multicolumn{1}{c}{Distribution Node ID}   \\\hline
\multicolumn{1}{c}{Commercial company}           & \multicolumn{1}{c}{7, 9, 11, 12, 32}            \\[-1ex] 
\multicolumn{1}{c}{Municipal department}         & \multicolumn{1}{c}{15, 18, 21, 31}             \\[-1ex] 
\multicolumn{1}{c}{Industrial company}           & \multicolumn{1}{c}{3, 5, 6, 8, 14, 23, 24, 25, 30} \\[-1ex]
\multicolumn{1}{c}{Hospital}                     & \multicolumn{1}{c}{4, 13}             \\[-1ex]
\multicolumn{1}{c}{Park}                         & \multicolumn{1}{c}{29}             \\[-1ex]
\multicolumn{1}{c}{Food service}                 & \multicolumn{1}{c}{10, 16, 17, 26, 28}         \\[-1ex] 
\multicolumn{1}{c}{Residential area}            & \multicolumn{1}{c}{2}           \\[-1ex] 
\multicolumn{1}{c}{Education}                   & \multicolumn{1}{c}{20}           \\[-1ex] 
\multicolumn{1}{c}{Integrated region}            & \multicolumn{1}{c}{19, 22}           \\[-1ex]
\multicolumn{1}{c}{Argriculture}                 & \multicolumn{1}{c}{27, 33}                   \\ \hline  \end{tabular}}
\end{table}

\subsection{Two-stage FCS-MCS joint model}

\begin{figure}[t]  
    \centering      \includegraphics[width=0.48\textwidth]{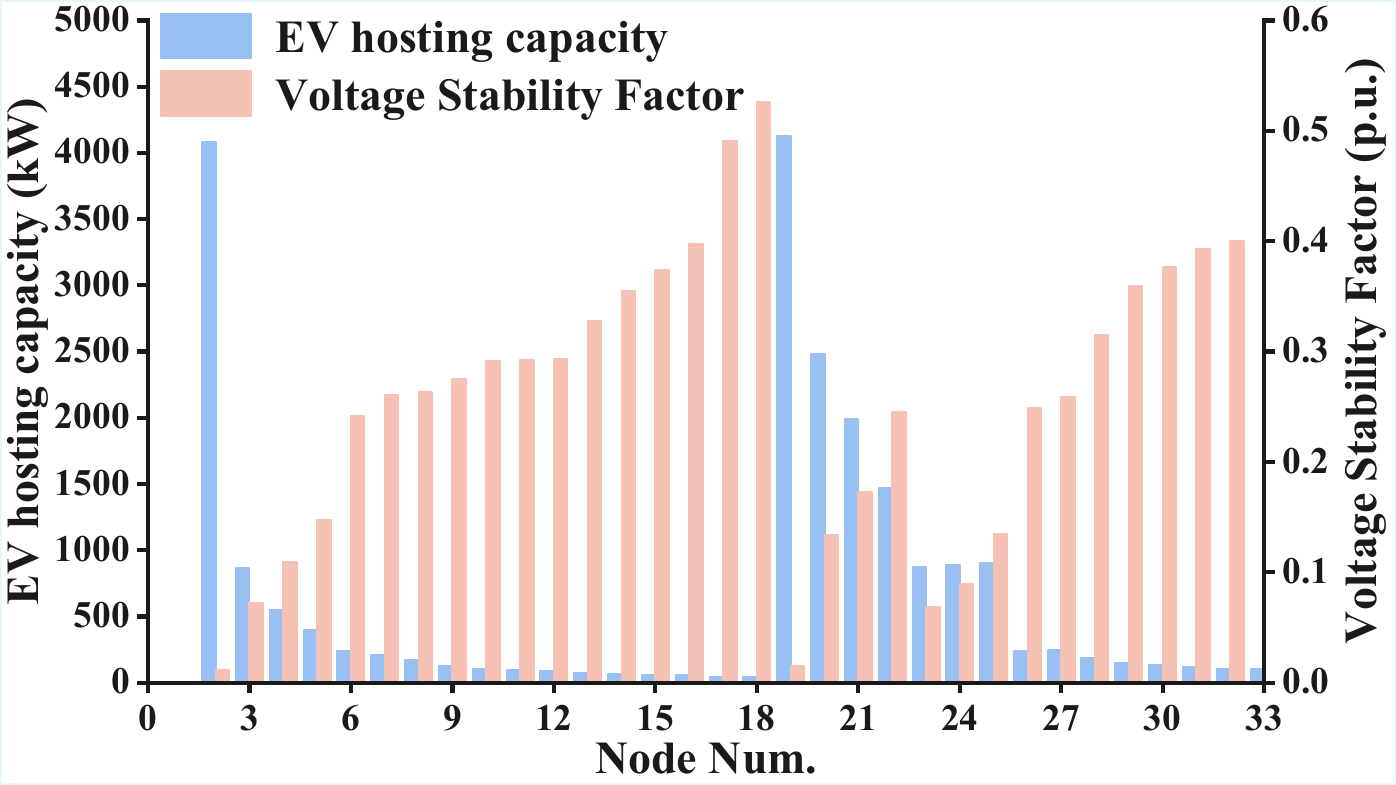}  
    \caption{EV-HC and VSF assessments in the 33-bus distribution network.}  
    
    \label{VSF}  
\end{figure}

Candidate locations for FCS deployment are determined using the assessment framework established in Stage 1.  VSF is solved by the forward and backward sweep algorithm with $d P_{EV,n}=0.01$. EV-HC is evaluated on Opendss with a stepsize of 5 kW and a safety margin of 0.85. The assessment results of VSF and EV-HC are depicted in Fig. \ref{VSF}. The nodes with the highest EV-HC are power nodes 19, 2, 20, 21, 22, while those with the lowest VSF are nodes 2, 19, 23, 3, 24. Due to the minimal differences in VSF values across the 33 nodes, EV-HC results are set as the primary selection criterion. Accordingly, nodes 19, 2, and 20 are selected as candidate locations in this study. 
    \label{hc}  

According to the joint model in Stage 2, three test cases based on the above candidate locations are conducted to evaluate the effectiveness of the proposed model. This section sets the background as flat time and uses average EV charging demand and average charging price to simulate common conditions. It is assumed that 80\% of EV users prefer FCS, while the remaining 20\% are equally willing to use either FCS or MCS. The convergence tolerances for both primal and dual residuals are set to $10^{-4}$, and the maximum number of iterations is set to 50. In this case study, the proposed model satisfies the convergence threshold within five iterations. The detailed results of three cases are presented in Table \ref{3cases}.   

\begin{table*}[]
\centering
\caption{Case results for the FCS-MCS joint model}
\vspace{-0.2cm}
\label{3cases}
\begin{tabular}{ccclllll} \hline
 Case&FCS location &
  \begin{tabular}[c]{@{}c@{}}FCS capacity\\[-1ex] (kW)\end{tabular} &
  \multicolumn{1}{c}{\begin{tabular}[c]{@{}c@{}}MCS locations \\[-1ex] and numbers\end{tabular}} &
  \multicolumn{1}{c}{\begin{tabular}[c]{@{}c@{}}FCS annual net  revenue \\[-1ex] (\$/year)\end{tabular}} &
  \multicolumn{1}{c}{\begin{tabular}[c]{@{}c@{}}MCS operation  cost \\[-1ex] (\$/h)\end{tabular}}  &
  \multicolumn{1}{c}{\begin{tabular}[c]{@{}c@{}}Total waiting time  \\[-1ex] cost at MCSs (\$)\end{tabular}} &
  \multicolumn{1}{c}{\begin{tabular}[c]{@{}c@{}}Total driving\\[-1ex]  distances (km)\end{tabular}} \\ \hline
 1&19 &
  1956.77 &
  \multicolumn{1}{c}{\begin{tabular}[c]{@{}c@{}}16 (3), 20 (4),\\[-1ex]23 (5), 25 (2)\end{tabular}} &
  \multicolumn{1}{c}{547846.7} &
  \multicolumn{1}{c}{12.6}  &
  \multicolumn{1}{c}{37.16} &
  \multicolumn{1}{c}{4480} \\[-1ex]
 2&2 &
  1962.88 &
  \multicolumn{1}{c}{\begin{tabular}[c]{@{}c@{}}12 (4), 20 (4),\\[-1ex]23 (5), 25 (2)\end{tabular}} &
  \multicolumn{1}{c}{551089.53} &
  \multicolumn{1}{c}{13.5}  &
  \multicolumn{1}{c}{41.15} &
  \multicolumn{1}{c}{3799} \\[-1ex]
 3&20 &
  1904.03 &
  \multicolumn{1}{c}{\begin{tabular}[c]{@{}c@{}}4 (5), 20 (4),\\[-1ex]23 (5), 25 (2)\end{tabular}} &
  \multicolumn{1}{c}{519840.72} &
  \multicolumn{1}{c}{14.4}  &
  \multicolumn{1}{c}{37.16} &
  \multicolumn{1}{c}{2937} \\ \hline \end{tabular}
\end{table*}
\vspace{-0.4cm}
\subsection{Flexible capacity strategy}
\begin{figure*}[!t]
\centering
\subfloat[Case 1]{\includegraphics[width=0.34\textwidth]{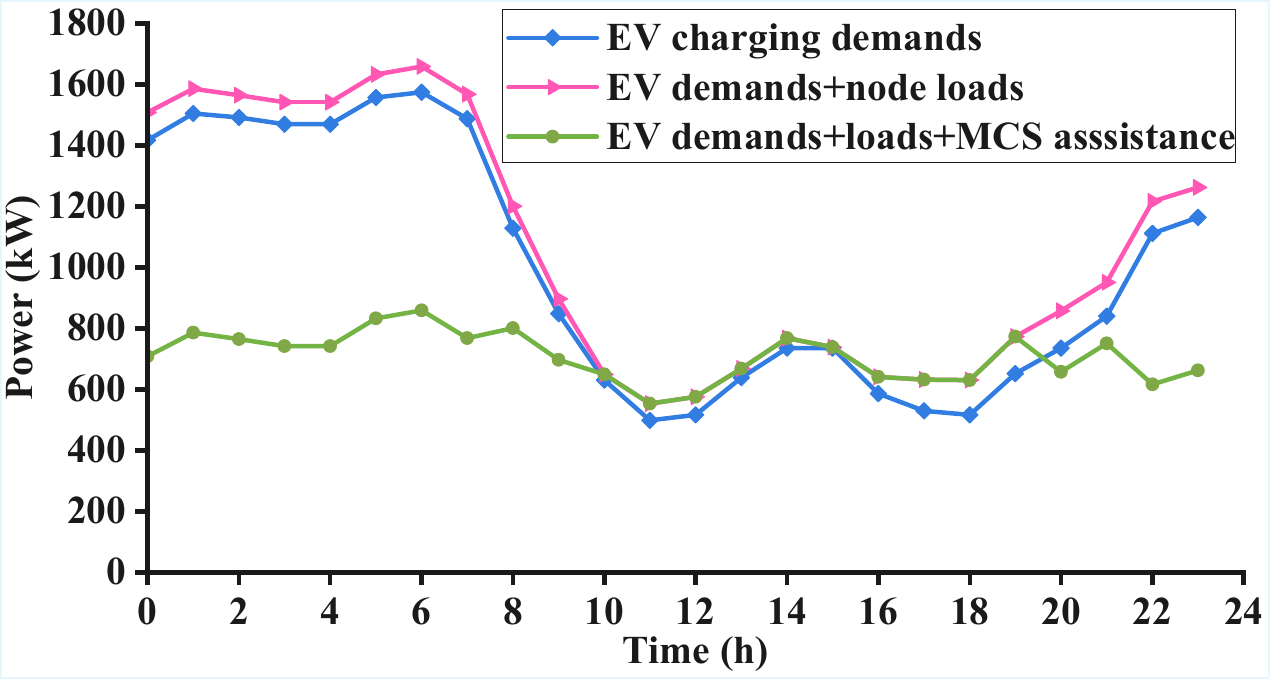}
\label{fig_first_case}}
\hspace{-5mm}
\subfloat[Case 2]{\includegraphics[width=0.34\textwidth]{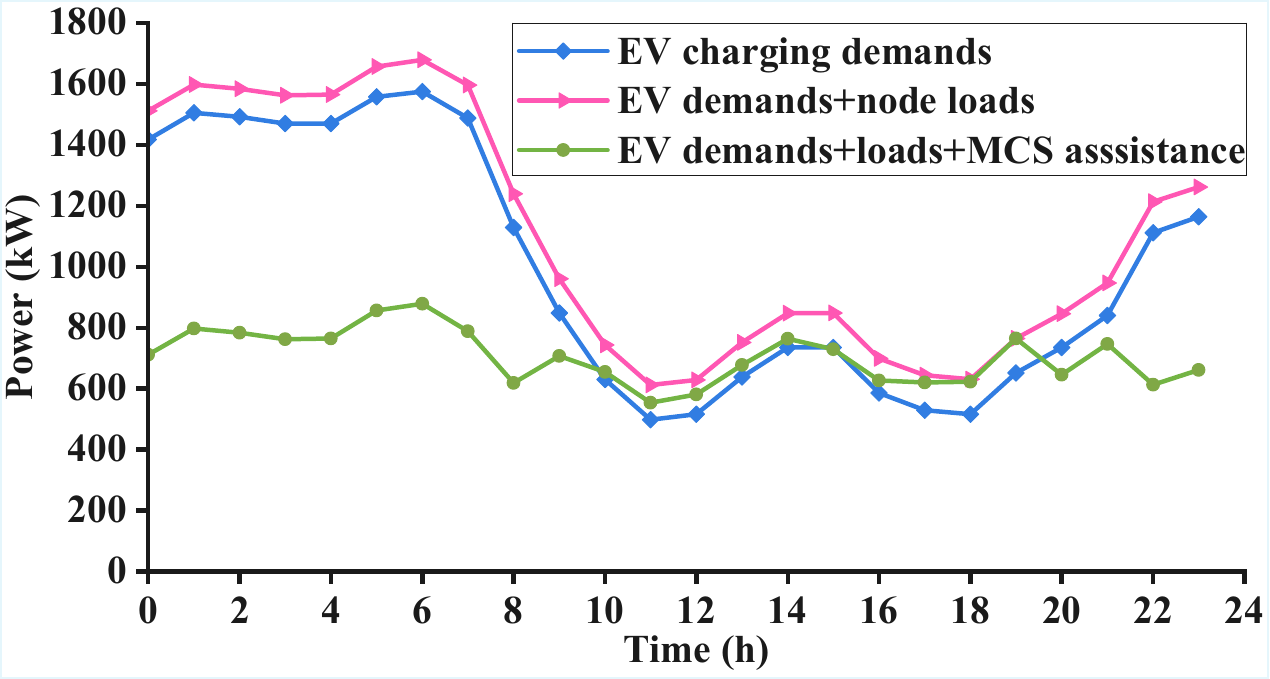}
\label{fig_second_case}}
\hspace{-5mm}
\subfloat[Case 3]{\includegraphics[width=0.34\textwidth]{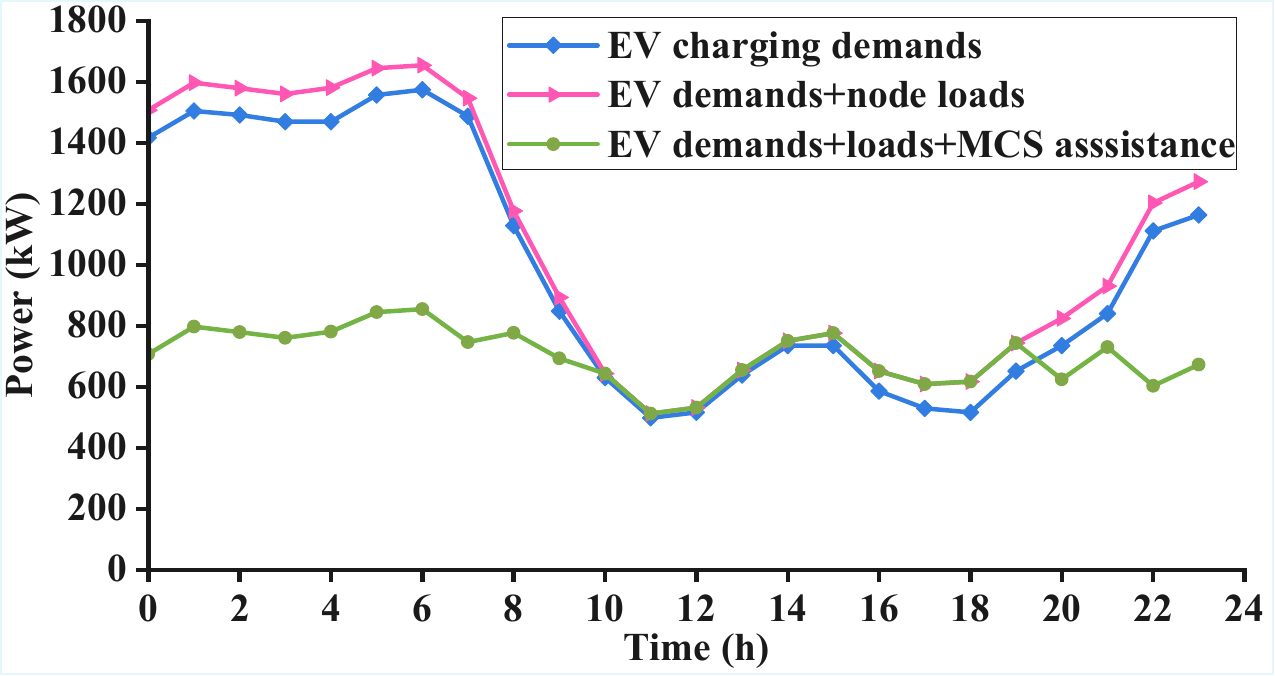}
\label{fig_third_case}}
\caption{The diagrams of load regulation through the proposed EMS.}
\label{figems}
\end{figure*}

As discussed previously, an energy management strategy is proposed to alleviate peak-time pressure and reduce capital investment while meeting EV charging demands. Table \ref{tabems} summarizes the flexible capacity across different timeframes of the three cases. The peak load regulation capability in this study refers specifically to the V2G functionality of MCS, which means the capacity of accommodating additional load during peak periods while smoothing the overall load curve. MCSs designated for V2G operation are configured to discharge their batteries from 95\% to 60\% state of charge (SoC). The effectiveness of the proposed EMS on load regulation is demonstrated in Fig. \ref{figems}. The node loads used in this part contain daily loads, PV generation, and user activities. In cases 1, 2, and 3,  1–6 MCSs are deployed per hour for V2G support during peak periods.  

\begin{table}[t]
\caption{Results for the flexible capacity strategy}
\label{tabems}
\vspace{-0.2cm}
\begin{tabular}{>{\centering\arraybackslash}p{0.05\linewidth}>{\centering\arraybackslash}p{0.20\linewidth}>{\centering\arraybackslash}p{0.10\linewidth}>{\centering\arraybackslash}p{0.10\linewidth}>{\centering\arraybackslash}p{0.25\linewidth}}
\hline
\multirow{2}{*}{Case} &
  \multirow{2}{*}{\begin{tabular}[c]{@{}c@{}}Capacity potential\\[-1ex] at original site\\[-1ex] (kW)\end{tabular}} &
  \multicolumn{2}{c}{EVCS flexible energy} &
  \multirow{2}{*}{\begin{tabular}[c]{@{}c@{}}Ability of regulating \\[-1ex] peak loads\end{tabular}} \\[-1ex]
  &         & \begin{tabular}[c]{@{}c@{}}FCS\\[-1ex] (kW)\end{tabular} & \begin{tabular}[c]{@{}c@{}}MCS\\[-1ex] (kWh)\end{tabular} &           \\\hline
1 & 2173.23 & 1956.77                                            & 8400                                                & 3.8 times \\[-1ex]
2 & 2122.12 & 1962.88                                            & 9000                                                & 4.1 times \\[-1ex]
3 & 580.97  & 1904.03                                            & 9600                                                & 4.3 times \\\hline
\end{tabular}
\end{table}
\vspace{-0.4cm}
\subsection{Comparison tests}
To evaluate the performance of the proposed two-stage FCS-MCS joint model, two comparison scenarios are conducted in this study. 
\subsubsection{Scenario 1}
In this scenario, the optimal locations for 5 FCSs are determined based on the p-median method. The selected coupling nodes are nodes 4, 5, 6, 7, and 8. In this scenario, without the capacity assessment in Stage 1, the node’s holding ability to EV loads is initially unknown. Therefore, a conservative planning method is applied, where EV charging loads are incrementally added in small steps on candidate locations, and iterations continue until the power constraint is violated. The voltage lower limit is set to 0.905 p.u. when the EV load is below 100 kW and 0.90 p.u. when it exceeds 100 kW. To allow a fair comparison with the proposed model, this scenario is configured to approximate its capacity. The step size is set to 25 kW, and the final planning capacities are 500 kW for coupling nodes 4, 5, 7, 8, and 50 kW for node 6.  
\subsubsection{Scenario 2}
\begin{figure}[t]  
    \centering      \includegraphics[width=0.36\textwidth]{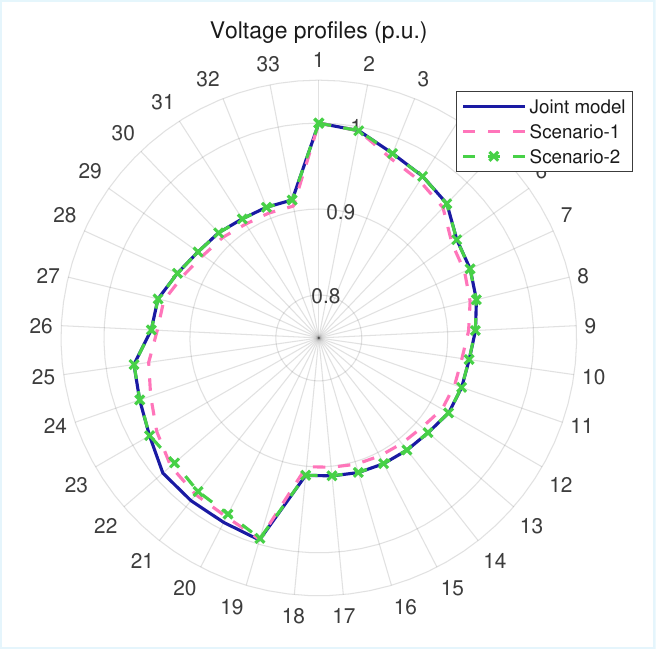}    
    \caption{Voltage profiles of IEEE 33-bus distribution network after charging station placement.}  
    \label{voltage}  
\end{figure}

In contrast, scenario 2 incorporates the assessment framework in Stage 1, where the EV-HC of each node is given. Nodes with the greatest EV-HC are selected for FCS placement in order to reduce operational pressure on the power system and decrease the cost of capacity expansion. Considering both service coverage and node hosting capacity, coupling nodes 1, 2, and 4 are selected as FCS locations. This scenario is also configured to approximate the capacity of the proposed model. The final planning capacities are 700 kW for each FCS. The basic investment cost is calculated with $C_{base}$ and $C_{inve}$. The basic results of the two comparison scenarios are presented in Table \ref{comparison}.  Fig. \ref{voltage} shows their impacts on voltage in the IEEE 33-node power system. 

\begin{table*}[]
\centering
\caption{Results of comparison tests}
\vspace{-0.2cm}
\label{comparison}
\begin{tabular}{ccllll} \hline
Scenario &
  \begin{tabular}[c]{@{}c@{}}Total fixed capacity\\[-1ex] (kW)\end{tabular} &
  \multicolumn{1}{c}{\begin{tabular}[c]{@{}c@{}}Basic investment cost \\[-1ex]  (\$)\end{tabular}} &
  \multicolumn{1}{c}{\begin{tabular}[c]{@{}c@{}}Flexible energy (short-term)\\[-1ex]  (kWh)\end{tabular}} &
  \multicolumn{1}{c}{\begin{tabular}[c]{@{}c@{}}Capacity expansion potential \\[-1ex] (kW)\end{tabular}} &
  \multicolumn{1}{c}{\begin{tabular}[c]{@{}c@{}}Total user driving distance \\[-1ex] (km)\end{tabular}} \\ \hline
Joint model          & 1962.88&\multicolumn{1}{c}{674,687}& \multicolumn{1}{c}{9000} & \multicolumn{1}{c}{2122.12} & \multicolumn{1}{c}{3799}  \\[-1ex]
Scenario-1           & 2050                 & \multicolumn{1}{c}{1,843,850} & \multicolumn{1}{c}{0}    & \multicolumn{1}{c}{2115}    & \multicolumn{1}{c}{700}  \\[-1ex]
Scenario-2           & 2100& \multicolumn{1}{c}{1,277,700}& \multicolumn{1}{c}{0}    & \multicolumn{1}{c}{7585}    & \multicolumn{1}{c}{3278} \\ \hline \end{tabular}
\end{table*}
\vspace{-0.2cm}
\vspace{-0.2cm}
\subsection{Discussion}
\label{} 
\subsubsection{Economic performance}
Compared with conventional planning scenarios commonly adopted in engineering practice, the proposed FCS-MCS joint planning model exhibits superior economic performance. Under conditions of approximate capacity and user service level, the joint planning model requires only one-third to one-half of the basic investment in comparison scenarios. In addition, the proposed model additionally provides 9000 kWh of flexible energy, demonstrating advantages in cost-efficiency and resource utilization. 
\subsubsection{Impact on power system operation}
Under the proposed joint planning model, the IEEE 33-node power distribution system operates reliably within parameter limits, and voltage margins remain adequate. Scenario 1, which features a distributed multi-point charging station layout, presents certain risks to voltage stability. Both Scenario 2 and the proposed model exhibit approximate stability performance, as illustrated in Fig. \ref{voltage}. Moreover, the proposed model also exhibits a strong capability to smooth peak load as shown in Fig. \ref{tabems}.
\subsubsection{Capacity expansion potential}
From the perspective of long-term scalability, expanding capacity at existing charging sites proves more cost-effective than constructing new stations. Although Scenario 2 offers the highest capacity expansion potential, it may lead to underutilization or resource waste. Both the proposed model and Scenario 1 provide comparable scalable capacity. The joint model achieves a better balance between capacity potential and initial investment. This makes it a more economical and adaptable solution for long-term development.
\vspace{-0.4cm}
\section{Conclusion}

\label{sec:6}
his paper proposes a two-stage joint planning model for FCSs and MCSs to transform the negative impacts from EV charging into positive effects. Optimizing EVCS location alone cannot fundamentally reduce the additional pressure introduced by EV charging, which merely identifies relatively better locations. Incorporating MCSs into the planning framework can provide flexible supplementary energy, improve the utilization rate of charging facilities, reduce investment, and enhance service quality. The proposed model first identifies candidate locations for FCSs based on EV-HC assessment. It then formulates a joint planning model for FCSs and MCSs, where a modified ADMM is employed to effectively coordinate different stakeholders. Furthermore, building on the above models, a flexible capacity planning strategy is proposed by considering capacity expansion potential, collaborative planning, and real-time operational management. This strategy can address the spatio-temporal uncertainties of charging loads and accommodate future charging demand growth, thereby alleviating operational pressure on the power grid. The case studies demonstrate that the proposed method provides economic planning plans for EVCS, while significantly improving the stability of power grid operation. Regarding future research directions, the impacts of user behavior and preference on EVCS joint planning need to be further explored to improve social significance.


\bibliographystyle{IEEEtran}
\bibliography{Bibliography.bib}





\vfill

\end{document}